\documentclass[twocolumn]{aastex63}
\usepackage{graphics,epsf}
\usepackage{amsmath}                
\usepackage{amsfonts}               
\usepackage{amssymb}                
\usepackage{epsfig}                 
\usepackage{appendix}
\usepackage{graphicx}
\usepackage{float}
\usepackage{color}
\usepackage{multirow}
\usepackage{colortbl}
\usepackage[para,online,flushleft]{threeparttable}

\hypersetup{citecolor=blue, 
            linkcolor=red, 
            menucolor=blue, 
            urlcolor=blue}  

\newcommand{\m}{{~\rm m}}
\newcommand{\km}{{~\rm km}}
\newcommand{\s}{{~\rm s}}
\newcommand{\h}{{~\rm h}}

\newcommand{\g}{{~\rm g}}
\newcommand{\G}{{~\rm G}}
\newcommand{\K}{{~\rm K}}

\newcommand{\yr}{{~\rm yr}}

\newcommand{\kpc}{{~\rm kpc}}

\begin{document}

\title{XRISM observations of the cooling flow cluster A2029 support heating by mixing }


\author{Noam Soker}
\affiliation{Department of Physics, Technion Israel Institute of Technology, Haifa, 3200003, Israel;  soker@physics.technion.ac.il}

\begin{abstract}
I argue that the mixing-heating mechanism of the intracluster medium (ICM) is compatible with the new observations by the X-ray telescope XRISM that show the dispersion velocity in the cooling flow cluster of galaxies A2029 to be $\sigma_v=169 \pm 10 \km \s^{-1}$. Past jets from the central supermassive black hole induced turbulence; the velocity dispersion value indicates that the jets were powerful, as expected in the mixing-heating mechanism. Although the kinetic energy of the ICM turbulence that XRISM finds is short of heating the ICM and counter radiative cooling, the turbulence is fast enough to mix the hot shocked jets' material with the ICM on time scales shorter than the radiative cooling time.  The support of the mixing-heating mechanism from determining the turbulent velocity I claim for A2029 is similar to the conclusion from the X-ray observations by the X-ray telescope Hitomi of the Perseus cluster. 
\end{abstract}
 

\section{Introduction} 
\label{sec:intro}

In cooling flows in clusters of galaxies and galaxies, the radiative cooling time of the hot gas, $\tau_{\rm cool}$, is shorter than the cluster age. The mass cooling rate to low temperatures, below X-ray emission, is lower than the hot gas mass divided by $\tau_{\rm cool}$ (not necessarily by a large factor, e.g., \citealt{Fabianetal2024}). The low mass cooling rate implies active galactic nucleus (AGN) jets heat the intracluster medium (ICM). A major open question is the processes that transfer the kinetic energy of the AGN jets to ICM thermal energy. Theoretical processes to transfer jets' energy to the ICM include (1) shock waves (e.g., \citealt{Randalletal2015}), (2) sound waves (e.g., \citealt{TangChurazov2018}), (3) internal waves  (e.g., \citealt{Zhangetal2018}), (4) turbulence in the ICM (e.g.,  \citealt{Zhuravlevaetal2018}), (5) uplifting gas from inner regions (e.g., \citealt{HuskoLacey2023}), (6) cosmic rays (e.g., \citealt{FujitaOhira2013}), and (7) heating by mixing hot bubble gas (thermal gas and/or cosmic rays) with the ICM (e.g., \citealt{GilkisSoker2012}).

Very recent observations by the X-ray telescope XRISM yield a field-integrated ICM velocity dispersion of $\sigma_v = 169 \pm 10 \km \s^{-1}$ in the cooling flow cluster A 2029 \citep{XRISM2025}. They find that if this velocity dispersion is of isotropic turbulence, the Mach number is $0.21$, and the non-thermal pressure fraction is $2 \%$. I here argue that this velocity dispersion is compatible with the expectation of the mixing-heating mechanism.

\section{Heating by mixing} 
\label{sec:HeatMixing}

\cite{HillelSoker2017} compared the Hitomi observed ICM velocity dispersion of the Perseus cluster of galaxies, $164 \pm 10 \km \s^{-1}$ in the region of $r=30 \kpc$ to $r=60 \kpc$ (\citealt{Hitomi2016}), with their three-dimensional hydrodynamical simulations of jet-inflated bubbles in cluster cooling flows. The simulations have similar velocity dispersions to the observed ones. 
These three-dimensional hydrodynamical simulations also show that heating by mixing the material of the hot shocked jets with the ICM heats the ICM much more efficiently than sound waves, shocks, and turbulence. \cite{HillelSoker2017} concluded that the Hitomi observations of Perseus support the mixing-heating mechanism. 
From the perspective of velocity dispersion, therefore, the observed value of $\sigma_v = 169 \pm 10 \km \s^{-1}$ in A2029 is also compatible with the expectation of the mixing-heating mechanism. 

One large difference between the Perseus and A2029 is that the jets in Perseus inflated large bubbles, while in A2029 did not. Simulations of narrow jets show that even when jets do not inflate large bubbles, they induce large vortices that can mix hot shocked-jet gas with the ICM in their vicinity (e.g., \citealt{RefaelovichSoker2012}). For the mixing-heating mechanism, it is of no significance that the kinetic energy of the turbulence in A2029 is very small, $\ll 10\%$ of the ICM thermal energy; the role of the turbulence is to mix hot shocked-jet gas with the ICM.  

\begin{figure}
\begin{center}
\includegraphics[trim=0cm 10.0cm 0cm 0cm, clip, scale=0.45]{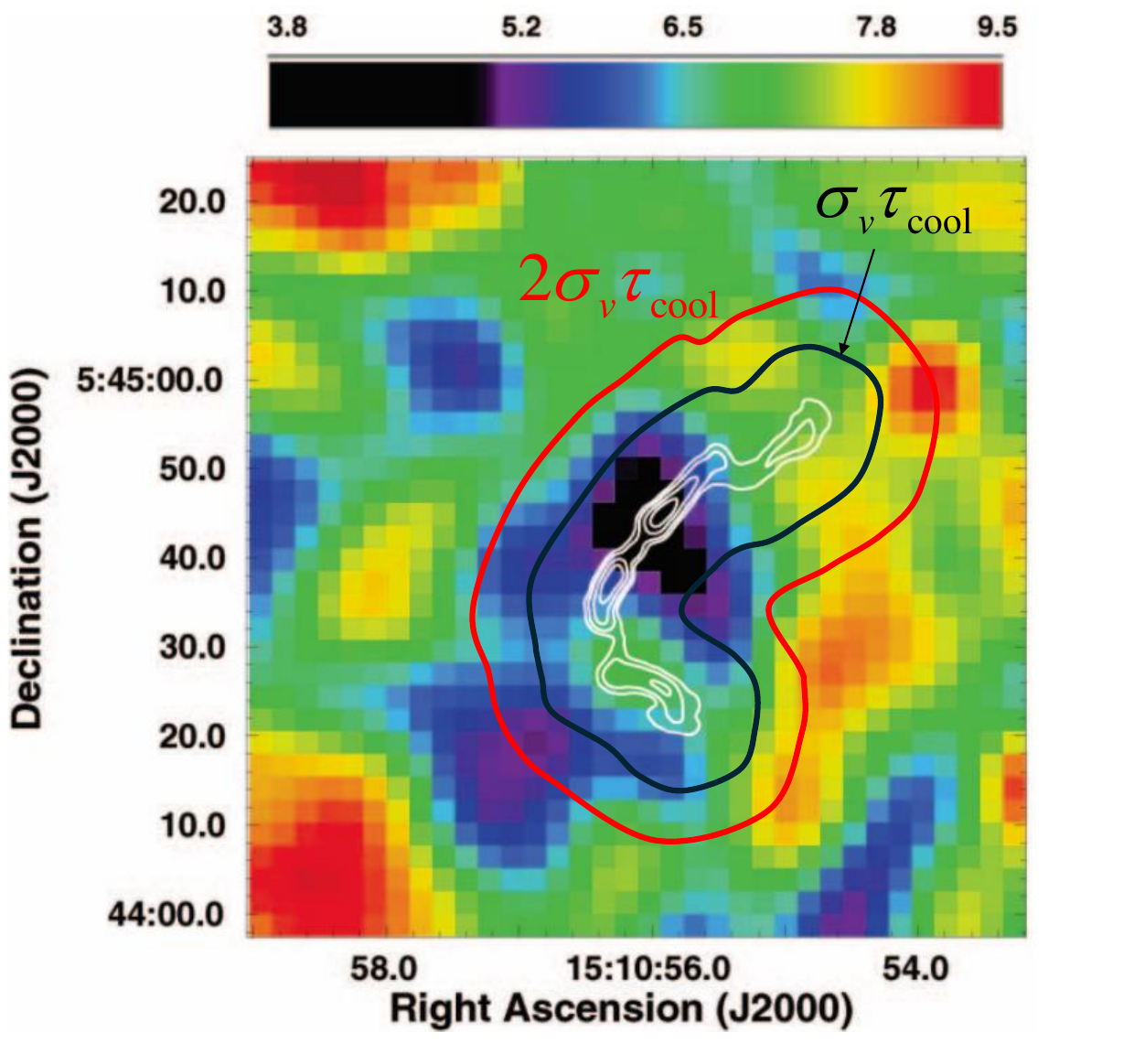} 
\caption{A temperature map with radio contours of A2029 adapted from \cite{Clarkeetal2004}. The temperature scale is by the color bar in keV; 1 arcsec corresponds to 1.45 kpc. 
I draw the approximate distance to which the turbulence can carry hot gas within one (black) and two (red) cooling times, where $\tau_{\rm cool}(0)=6 \times 10^7 \yr$.    
}
\label{fig:A2029}
\end{center}
\end{figure}

I take the mixing distance in time $\Delta t$ as $\sigma_v \Delta t$. Figure \ref{fig:A2029} presents the temperature map and radio contours from \cite{Clarkeetal2004}. I added two lines presenting the distances $\sigma_v \tau_{\rm cool}(0)$ (black) and $2 \sigma_v \tau_{\rm cool}(0)$ (red), where $\tau_{\rm cool}(0) = 5 \times 10^7 \yr$ is the cooling time of A2029 at the center \citep{Raffertyetal2008}. The distance the turbulence can carry energy at one central cooling time is $169 \km \s^{-1} \times  5 \times 10^7 \yr = 8.6 \kpc$. At a radius of $r=12 \kpc$ the cooling times is much longer $\tau_{\rm cool}(12) = 6 \times 10^8 \yr$  \citep{Raffertyetal2008}. To heat ICM zones away from the center, the turbulence has a time of $>\tau_{\rm cool}(0)$ to carry energy from the regions the jets shocked. 

\cite{HillelSoker2017}, compared their hydrodynamical simulations to the Hitomi observed velocity dispersion of the Perseus cluster and concluded that jets could excite the observed turbulence and are likely to do so. The same holds for A2029: the XRISM observed turbulence is compatible with that expected from past powerful AGN activity cycles.

\section{Summary} 
\label{sec:Summary}

XRISM's new high-quality observations allow us to explore the properties of the hot ICM medium of the cooling flow cluster A2029 and address the major question of the heating mechanism. The observed velocity dispersion implies that the turbulence kinetic energy is much smaller than the thermal energy of the ICM. However, the velocity dispersion value, like in the case of the Hitomi observations of the Perseus cluster, is compatible with turbulence excited by AGN jets that simulations yield \citep{HillelSoker2017}. The key point is that the role of turbulence is not to supply the energy to heat the ICM but rather to mix hot shocked gas in and near the jets with the ICM at larger distances. 

The new determination of the velocity dispersion allows us to estimate the distance to which the turbulence can mix hot gas within a given time in the frame of the mixing-heating mechanism. Figure \ref{fig:A2029} shows that the turbulence can carry hot gas to cover the cold center (black line) within the cooling time at the center. At those radii, the cooling time increases to larger values, and the mixing might operate longer to counter the radiative cooling.

A powerful AGN will be required to supply sufficient energy. The present AGN power of A2029 is lower than the radiative cooling rate (e.g., \citealt{Raffertyetal2006}). This might imply that some gas cools to lower temperatures and forms stars. However, \cite{Raffertyetal2006} noted that the central galaxy is red,  so it does not have a high star formation rate. According to the cooling flow feedback model, AGN activity cycles must have been much more powerful in the past. 

The new XRISM observations of A2029 are compatible with the feedback heating by the mixing-heating mechanism where the heating does not fully counter radiative cooling, implying the presence of a cooling flow. 


\label{lastpage}
\end{document}